\documentclass[final]{pasj01}
\usepackage{lscape} 

\begin{document} 
\Received{2018/07/20}%{yyyy/mm/dd}
\Accepted{2018/10/23}%{yyyy/mm/dd}
%\Published{yyyy/mm/dd}

\title{%
AKARI Mission Program:
Excavating Mass Loss History in Extended Dust Shells of Evolved Stars (MLHES)
I. Far-IR Photometry\footnote{Based on observations with AKARI, a JAXA project with the participation of ESA.}}

%%% begin:list of authors
% Do NOT capitalize all letters in "textsc".
\author{Toshiya \textsc{Ueta}\altaffilmark{1}}%
\email{toshiya.ueta@du.edu}
\author{Andrew J.\ \textsc{Torres}\altaffilmark{1}}
\author{Hideyuki \textsc{Izumiura}\altaffilmark{2}}
\author{Issei \textsc{Yamamura}\altaffilmark{3,4}}
\author{Satoshi \textsc{Takita}\altaffilmark{5}}
\author{Rachael L.\ \textsc{Tomasino}\altaffilmark{1}}

\altaffiltext{1}{%
Department of Physics and Astronomy,
University of Denver,
2112 E Wesley Ave.,
Denver, 80208, USA}
\altaffiltext{2}{%
%Okayama Astrophysical Observatory (OAO), 
%National Astronomical Observatory of Japan (NAOJ), 
%3037-5 Honjo, Kamogata, Asakuchi, Okayama, 719-0232, Japan}
%\altaffiltext{3}{%
Okayama Branch Office, Subaru Telescope, NAOJ, NINS
3037-5 Honjo, Kamogata, Asakuchi, Okayama, 719-0232, JAPAN}
\altaffiltext{3}{%
Institute of Space and Astronautical Science, 
JAXA, 
3-1-1 Yoshinodai, Chuo-ku, 
Sagamihara, Kanagawa, 252-5210, Japan}
\altaffiltext{4}{%
Department of Space and Astronautical Science, SOKENDAI,
3-1-1 Yoshinodai, Chuo-ku, 
Sagamihara, Kanagawa, 252-5210, Japan}
\altaffiltext{5}{%
National Astronomical Observatory of Japan, 
2-21-1 Osawa, Mitaka, Tokyo 181-8588, Japan}

%%% end:list of authors

%% `\KeyWords{}' always has to be placed before `\maketitle'.
\KeyWords{%
surveys --
circumstellar matter --
stars: mass-loss --
stars: AGB and post-AGB --
%stars: winds, outflows --
%dust --
infrared: stars} %Do NOT move this preamble from here!

\maketitle

\begin{abstract}
We performed a far-IR imaging survey of the circumstellar dust shells of 144 evolved stars as a mission programme of the AKARI infrared astronomical satellite using the Far-Infrared Surveyor (FIS) instrument. 
With this survey, we deliver far-IR surface brightness distributions of roughly 10$^{\prime}$ $\times$ 40$^{\prime}$ or 10$^{\prime}$ $\times$ 20$^{\prime}$ areas of the sky around the target evolved stars in the four FIS bands at 65, 90, 140, and 160\,$\micron$. 
Our objectives are to characterize the far-IR surface brightness distributions of the cold dust component in the circumstellar dust shells, from which we derive the amount of cold dust grains as low as 20\,K and empirically establish the history of the early mass loss history. 
In this first installment of the series, we introduce the project and its aims, describe the observations, data reduction, and surface brightness correction process, and present the entire data set along with the results of integrated photometry measurements (i.e., the central source and circumstellar dust shell altogether). 
We find that 
(1) far-IR emission is detected from all but one object at the spatial resolution about 30$^{\prime\prime}$ - 50$^{\prime\prime}$ in the corresponding bands, 
(2) roughly 60 -- 70\,\% of the target sources show some extension, 
(3) previously unresolved nearby objects in the far-IR are now resolved around 28 target sources, 
(4) the results of photometry measurements are reasonable with respect to the entries in the AKARI/FIS Bright Source Catalogue, despite the fact that the targets are assumed to be point-sources when catalogue flux densities were computed, and 
(5) an IR two-color diagram would place the target sources in a roughly linear distribution that may correlate with the age of the circumstellar dust shell and can potentially be used to identify which targets are more extended than others.
\end{abstract}

\section{Introduction} \label{sec:intro}

Low to intermediate initial mass stars (of 0.8--8\,M$_{\odot}$) experience copious mass loss at the rate of $10^{-8}$ to $10^{-4}$\,M$_{\odot}$\,yr$^{-1}$ in the late stages of evolution, especially during the asymptotic giant branch (AGB) phase (cf., \cite{herwig2005}, \cite{iben2012}, \cite{karakas2014}). 
This stellar mass loss is expected to be induced by specific physical conditions at the stellar surface as a consequence of the internal evolution, while the course of the internal evolution is thought to be influenced by this mass loss that impacts the physical conditions at the stellar surface.
For example, mass loss may show temporal variations because of the alternative burning of hydrogen and helium in distinct layers via a mechanism called ``thermal pulses" \citep{paczynski1971} during the AGB phase, while a sudden, drastic increase of mass loss -- the so-called ``superwind" \citep{renzini1981} -- near the end of the AGB phase can remove a significant portion of the surface layer from the star, forcing the star to terminate the AGB evolution.
The AGB mass loss, therefore, determines the fate of AGB stars by controlling the highest luminosity that they can attain and duration of the AGB phase, which would proportionally increase the lower limit of the zero-age main-sequence (ZAMS) mass that results in a supernova (c.f., \cite{weidemann1990}).
However, no theoretical description of mass loss has ever been derived fully from first principles, despite that identifying the exact mechanisms of mass loss in these cool stars has been and still is a long-standing problem in the stellar evolution. 

During the AGB phase, mass-loss ejecta from the central star form a circumstellar dust shell (CDS)\footnote{Indeed, a unique set of information, especially on the velocity and chemistry, of the mass-loss history can be derived from the gas component of the circumstellar gas shells (CGS).  Readers who are also interested in following the latest developments in the investigation of the gas component of the AGB circumstellar shells are encouraged to refer to recent reviews such as the one by \citet{ho18}.}. 
When mass loss is terminated at the end of the AGB phase, the CDS becomes physically detached from the central star and starts to drift away. 
At this point, the newly-formed post-AGB CDSs appear to have already developed largely axisymmetric structures (cf., \cite{meixner1999,ueta2000,meixner2002,sahai2007,siodmiak2008}).
Thus, the AGB mass loss affects not only the internal evolution but also the external evolution. 
While an ample amount of effort was made to confirm that the post-AGB CDSs do possess at least some axisymmetric structure, still lacking is the direct observational evidence for the structure development in the AGB CDSs.
In particular, the temporal evolution of the CDS structure development needs to be constrained by observations.
Hence, a thorough observational study of the AGB CDSs must be done to understand the earlier AGB mass loss history pertaining to the CDS structure formation. 

To follow the early AGB mass loss history observationally, one must employ far-IR thermal emission arising from the colder dust grains in the outermost regions of the CDSs as a primary tracer. 
While there is also molecular gas emission in the inner regions of the CDSs, molecules tend to be photo-dissociated by the interstellar radiation field (ISRF) in the outermost parts of the CDSs (e.g., \cite{meixner1998}). 
The presence of the ISRF does a favor for our purposes: (1) the ISRF would warm up dust grains in the outermost regions of the CDS at about $20 - 40$\,K (e.g., \cite{gillett1986,young1993a}), and hence, would enhance, however slightly, the detectability of the CDSs in the far-IR, and (2) the CDSs are likely more extended than the molecular shells because of photo-dissociation by the ISRF, and possibly, by dust/gas drift.
Thus, dust grains in these regions would not suffer from cooling by molecules.

Here, we present the far-IR imaging survey of evolved star CDSs at various evolutionary stages from the tip of the first giant branch to the planetary nebula (PN) phase including AGB and post-AGB phases, conducted using the AKARI infrared (IR) astronomical satellite and the Far-IR Surveyor (FIS) instrument.
Our primary goals of this survey are to 
(1) detect far-IR emission from cold dust grains that reside in the CDSs of evolved stars,
(2) perform photometry of the detected sources, and
(3) establish the density distribution in the CDSs from the detected far-IR surface brightness distributions for the CDSs without the influence of the central star and reconstruct the AGB mass loss history in detail, and
(3) quantify parameters concerning the AGB mass loss and CDS structures.
In this first installment of the series, we introduce the project, describe the observations, data reduction, and surface brightness correction process, and present far-IR images for the entire sample of 144 targets along with the results of integrated photometry measurements made for the entire object (i.e., the central star and CDS altogether).
Separation of the CDS and stellar emission via PSF removal and subsequent characterization of the CDS emission will be dealt with in the second installment of the series (Torres et al.\ {\sl in preparation}).

\section{Previous Far-IR Studies of the Evolved Star CDSs}

The existence of very extended CDSs was originally recognized by individual spatial investigations by, for example, \citet{hacking1985}, \citet{gillett1986}, and \citet{stencel1988}, using IRAS \citep{iras}.
Alternatively, \citet{willems1988} and \citet{chan1988} indicated the presence of the detached CDSs to explain the distribution of cool evolved stars on the IRAS 12-25-60\,$\micron$ two-color diagram.
Meanwhile, 
\citet{young1993a} and \citet{young1993b}
%Young et al.\ (1993a; 1993b) 
were the first to fully exploit the potential of IRAS survey scan data in the far-IR, fitting the scan data passing over a target object by a simple CDS model.
However, due to insufficient spatial resolutions, it was not possible to examine the internal structures of the extended CDSs around evolved stars until later when an advanced image processing technique was introduced (e.g., \cite{waters1994,izumiura1997,hashimoto1998}). 
Also, IRAS was not quite suited to probe the coldest dust emission from the evolved star CDSs as its reddest band was 100\,$\micron$.
Then, ISO \citep{iso} was used to observe a small sample (about 10) of evolved stars. 
Despite ISO's small sky coverage and large pixels, extended CDSs were detected and their structures were studied for both O-rich and C-rich AGB stars (e.g., \cite{izumiura1996,hi1998}).

Although observations made with IRAS and ISO were successful, observational evidence was still lacking if we strive for the level of understanding that would allow more detailed characterization of mass loss and CDS structure formation during the AGB phase and beyond. 
In particular, detector sensitivities had to be improved to detect much weaker thermal dust emission, which may arise from a weakly mass-losing stars or the outermost regions of very extended CDSs. 
The former is necessary to learn mass loss experienced by lower progenitor-mass stars, while the latter is essential to investigate even earlier phases of mass loss.

At the turn of the 21$^{\rm st}$ century, there was a {\it renaissance} of far-IR astronomy with the coming of the next generation of space telescopes launched successively -- 
the Spitzer Space Telescope (Spitzer; \cite{spitzer}) in 2003 by the National Aeronautics and Space Administration, 
the AKARI IR astronomical satellite (AKARI; \cite{akari}) in 2006 by the Institute of Space and Aeronautical Science (ISAS) of the Japan Aerospace Exploration Agency (JAXA), and 
the Herschel Space Observatory (Herschel; \cite{herschel}) in 2009 by the European Space Agency.
These new opportunities permitted us to probe CDSs at better sensitivities and spatial resolution than the preceding studies with IRAS and ISO.

However, observations of the extended CDS around evolved stars by Spitzer was met by a surprise. 
The bow-shock-like interface structure between the CDS and the interstellar medium (ISM) of $\sim 200^{\prime\prime}$ radius was detected in the far-IR around the AGB star, R Hya \citep{ueta2006}, as part of the MIPS IR Imaging of AGB Dust shells (MIRIAD; PI: A.\,K.\,Speck) program.
The detected structure was very much similar to the one found around Betelgeuse and other high-mass stars found earlier by IRAS, albeit smaller \citep{stencel1988,nc1997}. 
It was a surprise because the CDS-ISM interface structure was not expected around AGB stars, simply because the stellar wind velocity of AGB stars (of $10-20$\,km\,s$^{-1}$) was in general considered too low to cause shocks and active interactions at the CDS-ISM interface. 
As it turned out, what matters was the relative velocity of the stellar wind with respect to the local ISM. 
Hence, the CDS-ISM interactions can happen even for AGB stars if there is sufficient relative motion between the star and the local ISM \citep{ueta2011}.
In fact, Herschel followed suit to discover the CDS-ISM interface structures in the far-IR in a significant fraction of the AGB stars and red supergiants in the sample (50 out of 78, or 63\,\%; \cite{cox2012}) observed as part of the Mass-loss of Evolved StarS (MESS; PI: M.\,A.\,T.\,Groenewegen) guaranteed time key program \citep{mess}.

Meanwhile, a number of programs were executed
to address the issue of AGB mass loss by probing the CDS density distributions in the far-IR,
aiming to capture the onset of the axisymmetric structure development. 
Besides MIRIAD and MESS mentioned above,
these programs include,
%MIRIAD (PI: A.\ K.\ Speck; \cite{ueta2006,izumiura11}), 
COASTING (PI: M.\ Morris; \cite{do2007}), and 
Spitzer-MLHES (PI: T.\ Ueta; \cite{ueta2010})
with Spitzer,
MLHES (PI: I.\ Yamamura; \cite{ueta2007,izumiura2009}), and %,ueta2010,izumiura2011,ueta2011,ueta2017}) and
FISPN (PI: P.\ Garc{\'{\i}}a-Lario; \cite{cox2011})
with AKARI,
and 
%MESS (PI: M.\ A.\ T.\ Groenewegen; e.g., \cite{mess}) and
HerPlaNS (Herschel Planetary Nebula Survey; PI: T.\ Ueta; \cite{ueta2017,otsuka2017})
with Herschel.

Unfortunately, mapping extended CDSs with Spitzer turned out to be challenging because 
(1) a significant portion of the MIPS 70 and 160\,$\micron$ arrays were lost during launch and 
(2) the central star had to be avoided to prevent the MIPS 24\,$\micron$ array from saturating 
(even when the 24\,$\micron$ data were not needed).
Nevertheless, both of the MIRIAD and Spitzer-MLHES programs yielded a certain amount of results:
besides the R Hya discovery of the CDS-ISM interface \citep{ueta2006}, another peculiar CDS-ISM interface case of R Cas was studied \citep{ueta2010}.
However, because Spitzer CDS maps usually lack the central $50^{\prime\prime} \times 100^{\prime\prime}$ region, the latest mass loss history cannot be learned effectively from these maps.

The MESS program was a major survey of the evolved star CDSs with Herschel.
The power of its unprecedented spatial resolution was seen from the detection of wakes due to instabilities in the CDS-ISM interface structures around X Her and TX Psc \citep{jorissen2011} and the CDS-ISM interface structure around Betelgeuse resolved into three separate arcs \citep{decin2012},
besides the discovery of a number of CDS-ISM interaction cases \citep{cox2012}.
As for more genuine CDS cases, \citet{kerschbaum2010} detected a detached circular CDS almost coincident with the previously known CO circular shell around AQ And, U Ant, and TT Cyg.
Moreover, following the discovery of the spiral structures due to mass loss modulations of the central binary system in CO \citep{maercker2012}, \citet{mayer2013,mayer2014} found internal spiral structures in the far-IR CDS characteristic to mass loss modulations of the central binary system.

Some of the AKARI maps were also already presented in the context of the CDS-ISM interface structure \citep{ueta2008,ueta2010}.
\citet{izumiura2011} revealed a detached CDS after removing the point-spread function (PSF) effects due to the central star by subtracting a scaled image of a reference point source and interpreted the deduced radial profiles as caused by 
(1) the temporal enhancement of mass loss due to thermal pulse and the subsequent two-wind interactions or
(2) the reverse/termination shock of the stellar wind bounced back from the CDS-ISM interface.
While the former scenario was likely and preferred, the latter scenario was not completely refuted by the presence of the cometary tail structure seen (Fig.\,8 in \cite{izumiura2011}) and the marginal image quality at the preliminary stage of the data reduction.

\section{Observations} \label{sec:tech}

\subsection{AKARI Infrared Astronomical Satellite}

AKARI (formerly ASTRO-F) was the Japanese IR space mission launched on 2006 February 21 (UT). 
The mission goals of AKARI are to 
(1) perform a high spatial resolution all-sky survey in six IR bands from 9 to 160\,$\micron$, and 
(2) conduct pointed observations of specific targets to obtain deeper images and spectroscopic data from 2 to 180\,$\micron$. 
AKARI carried out its 550-day cryogenic mission until it exhausted its liquid helium on 2007 August 26, and continued its post-cryogenic mission in the near-IR until the satellite was finally turned off on 2011 November 24.

The Far-IR Surveyor (FIS: \cite{kawada2007}) was one of the two instruments on-board AKARI, 
covering the wavelength range of 50 to 180\,$\micron$ with two sets of Ge:Ga arrays, 
the Short Wavelength (SW) detector in the 
%N60 (at 65\,$\micron$ with $\Delta\lambda=21.7$\,$\micron$) and 
%WIDE-S (at 90\,$\micron$ with $\Delta\lambda=37.9$\,$\micron$) bands 
N60 (50--80\,$\mu$m, the reference wavelength $\lambda_\textrm{ref} = 65$\,$\mu$m) and
WIDE-S (60--110\,$\mu$m, $\lambda_\textrm{ref} = 90$\,$\mu$m) bands
\citep{fujiwara2003} and 
the Long Wavelength (LW) detector in the 
%WIDE-L (at 140\,$\micron$ with $\Delta\lambda=52.4$\,$\micron$) and 
%N160 (at 160\,$\micron$ with $\Delta\lambda=34.1$\,$\micron$) bands
WIDE-L (110--180\,$\mu$m, $\lambda_\textrm{ref} = 140$\,$\mu$m) and
N160 (140--180\,$\mu$m, $\lambda_\textrm{ref} = 160$\,$\mu$m) bands
\citep{doi2002}. 
During all-sky survey observations, the sky was swept at $3\farcm6$\,s$^{-1}$ covering more than 98\% of the entire sky (the all-sky scan mode; \cite{doi2015,takita2015}). 
During pointed observations, on the other hand, intended targets were scanned about $\times 20$ slower 
at 8 or 15$^{\prime\prime}$\,s$^{-1}$ to achieve one to two orders of magnitude better sensitivity than the all-sky survey observations (the slow-scan mode; \cite{shirahata2009}).

As we will demonstrate below, AKARI slow-scan maps achieve the 1-$\sigma$ sensitivity of less than 1\,MJy\,sr$^{-1}$, thanks to the slow scan-mapping speed, its cooled mirror, and the marginal spatial resolution (as opposed to Herschel maps that went as deep as $\gtrsim 1$\,MJy\,sr$^{-1}$ 1-$\sigma$ sensitivities; cf.\ \cite{mess,ueta2014}).
Hence, AKARI CDS maps are by far the most sensitive far-IR images of evolved star CDSs ever produced.

\subsection{Observing Strategy}

Our primary aim is to establish the AGB mass loss history observationally to enhance our understanding of mass loss and the CDS structure formation. 
To trace the history of AGB mass loss over the last $10^{5}$ years, it is necessary to examine the CDS out to 1.5\,pc from the central star assuming the shell expansion velocity of 15\,km\,s$^{-1}$. 
This means that the apparent size of the CDSs is on the order of $100^{\prime\prime}$ at typical distances of evolved stars ($\sim 1$\,kpc).
Meanwhile, we also need to cover the sufficient amount of blank sky to assure reliable background subtraction for the clear detection of CDSs. 
Furthermore, we must examine the CDS structures in 2-D maps to distinguish between a real extension of dust emission and a local cirrus that mimics a dust shell in a 1-D scan. 
2-D mapping is also important to study non-spherical morphologies expected in the CDSs around AGB and post-AGB stars.
Therefore, the use of 2-D mapping in our investigation is essential, and thus, we employ the Astronomical Observation Template (AOT) FIS01, which is designed for photometry observations of compact sources in the slow-scan mode, to make scan maps of $10^{\prime} \times 20^{\prime}$ (width $\times$ length) typically in the SW bands. 

This AOT performs two round-trip scans along the in-scan direction in the vicinity of the target source during the $\sim$\,10\,min window of pointed observation in the 90-min sun-synchronous orbit.
Between two round-trip scans, there is a small offset along the cross-scan direction to secure coverage of the target source by different detector elements. 
In the end, this AOT provides four independent redundant scans of the target to be combined into a single map, assuring the reliability of the resulting map.
For the actual execution of the AOT, we use
a reset interval of 0.5, 1.0, or 2.0\,sec,
the scan speed of 8 or 15$^{\prime\prime}$\,s$^{-1}$,
and the cross-scan offset of 70 or 240$^{\prime\prime}$ depending on the expected  
surface brightness and size of the target.

\subsection{Target Selection}

Our understanding of mass loss and the CDS structure formation can only be enhanced by collecting data from a large number of sources at various evolutionary stages (i.e., at the AGB, post-AGB/proto-PN, and PN phases), of different pulsation types (i.e., of Lb, SR, and Mira), and of different chemical types (i.e., C-, M-, and S-type). 
For the AKARI MLHES mission program, we establish first a volume limited sample of evolved stars before the launch of AKARI by the following procedure:
\begin{description}
\item[AGB stars]  
1) we extract known Lb, SR(a,b,c), and Mira type variables from the IRAS point source catalog, 
2) we calculate the bolometric flux for each source using the 12\,$\micron$ flux density and a bolometric correction based on the mid-IR color by $\log(F_{25}/F_{12})$, 
3) we estimate the distance to these sources by assuming a bolometric luminosity of 2,500\,L$_{\odot}$ for all sources and extract those closer than 500\,pc. 
Note that the luminosity is set conservatively somewhat smaller than the average value of 3,000\,L$_{\odot}$ for AGB stars (e.g., 2,600\,L$_{\odot}$ by \cite{habing1985}; 3,000\,L$_{\odot}$ by \cite{knauer2001}; 3,500\,L$_{\odot}$ by \cite{jackson2002}).

\item[OH/IR stars] 
1) we extract IRAS point sources which have mid-IR colors typical for OH/IR stars, 
2) we calculate the bolometric flux for each source using the 12\,$\micron$ flux density and a bolometric correction based on the mid-IR color, 
3) we estimate the distance to these sources by assuming a bolometric luminosity of 5,500\,L$_{\odot}$ for all sources and extract those closer than 1,000\,pc. 
Note that the distance limit for OH/IR stars is larger because the space density of OH/IR stars is very low and we do not find any of them within 500\,pc.

\item[Post-AGB stars/Proto-PNe] 
1) we examine literature for known post-AGB candidates (e.g., \cite{szczerba2007}), 
2) we extract sources which were found to be extended in the far-IR (e.g., \cite{speck2000}).

\item[PNe] 
1) we extract known PNe in the IRAS point source catalog (e.g., \cite{acker1992}), 
2) we select those with galactic latitude greater than $20^{\circ}$,
3) we extract those that are bright in the IR using the Innsubruck Data Base of Galactic Planetary Nebulae \citep{kk1997}.

\item[Extended Sources] 
We also include evolved objects which were previously found to be extended in the IRAS/ISO far-IR bands in the literature.
\end{description}

We then check the AKARI visibility of the selected targets and exclude those with low visibilities. %(of ``0" and ``1").
Among the resulting sample, we give the highest priority to those that
(1) show extension in the previous IRAS survey scan data (e.g., \cite{young1993a,young1993b}), 
(2) are reported to possess an extended CDS in the ISO studies (e.g., \cite{izumiura1996,hi1998}), 
and 
(3) are found to be extended in our unpublished studies using the High Resolution IRAS images and ADDSCAN data.
We also take into account the critical observing conditions/constraints such as saturation, sensitivity, map size, and detection reliability. 
Sources located in high far-IR background regions are removed from the target list:
we set the limiting threshold as 20\,MJy\,sr$^{-1}$ at 100\,$\micron$. 
However, the last criterion greatly reduces the number of promising candidates.
Last but not least, the target size is practically limited by the fact that we are awarded with 150 pointings, of course.

\subsection{Observations}

The science operation of AKARI, consisting of large-area survey programs (LSs), mission programs (MPs), and open-time programs (OTs), began on 2006 May 8. 
The all-sky survey was prioritised during Phase 1, which lasted until 2006 November 10. 
In addition to the all-sky survey, surveys of the North Ecliptic Pole and Large Magellanic Cloud were conducted as the LS program. 
Most of the MPs and OTs, including our observations, were executed during Phase 2, which followed Phase 1 and continued until the end of AKARI's cold campaign.
In the end, 149 pointed observations are made by AKARI under the MLHES MP between 2006 September 9 and 2007 August 20, of which 144 are FIS scans in the far-IR and five are IRC mapping in the mid-IR.
Some of the IRC maps have already been presented \citep{arimatsu2011}. 
Table \ref{tab:targets} summarizes the 144 FIS targets and their basic characteristics, 
together with a log of observations including
target name (usually the target's IRAS designation),
alternative/more common name, 
equatorial coordinates (J2000), 
Simbad object type,
variability type,
spectral type,
date of observations,
time of observations,
observation ID (a unique number to specify an AKARI data set),
and 
AOT parameters.
% 20 AGB, 
% 31 C-star, 
% 8 S-star, 
% 30 LPV, 
% 24 Mira, 
%
% 3 OH/IR, 
%
% 4 post-AGB, 
%
% 17 PNe, 
%
% 1 RCrB, 
% 3 supergiants, 
% 1 symbiotic, 
% 1 nova - do we include it?
% 1 SN 

\section{Data Reduction}

The AKARI/FIS slow-scan mapping data are archived in the Data ARchives and Transmission System (DARTS) maintained by ISAS/JAXA\footnote{http://darts.isas.jaxa.jp/astro/akari/}
in the time-series data (TSD) format.
After the archived TSD sets are downloaded from DARTS, they need to be processed into co-added far-IR maps with a map-making tool. 
We employ the second-generation data reduction package, FIS AKARI Slow-scan Tool (FAST: \cite{ikeda2012}; also briefly described by \cite{ueta2017}).
While FAST is similar to the first-generation data reduction package, FIS Slow-Scan data analysis Toolkit (SS-Tool: \cite{matsuura2007}), FAST can perform superior glitch and calibration lamp after-effect removal \citep{suzuki2008}. 

During the AOT FIS01 scan sequence in orbit, calibration measurements are taken five times before and after each of the four parallel scan legs with the shutter closed \citep{kawada2007}.
The dark current and calibration lamp signals are monitored during each of these five calibration exposures to follow the time-varying instrument responsivity. 
On the other hand, self flat-field measurements, in which a flat frame is constructed to correct for the pixel-to-pixel detector responsivity variations, are made with the shutter opened by exposing the detector to the ``flat'' sky during the calibration sequences at the beginning and end of the entire scan sequence (pre-cal and post-cal) while the telescope transitions between the all-sky survey mode and the slow-scan pointed observation mode \citep{matsuura2007}.

While SS-Tool is set to use all of the dark subtraction, detector responsivity time variation correction, and the pre-cal flat-fielding, FAST allows users to determine whether or not particular calibration measurements are used in the data processing with the help of the GUI, which visualizes the calibration measurements in the TSD. 
This flexible selection of calibration measurements with FAST greatly enhances the effectiveness of the corrections and improves the resulting data quality.
This is especially true because it is now possible to discard calibration data that are compromised by anomalies. 
These improvements available in FAST results in cleaner final co-added maps than those made by SS-Tool.

Yet another difference between SS-Tool and FAST is how photon energy is assumed to be distributed over the FIS arrays for each photon hit in respective map-making processes. 
With SS-Tool, photon energy is always distributed uniformly within the pre-determined beam size (i.e., a boxcar/pillbox function) of $40^{\prime\prime}$ and $60^{\prime\prime}$ for the SW and LW band, respectively.
With FAST, however, a variety of gridding convolution functions such as Gaussian and sinc functions are available to users \citep{ikeda2012}. 
For the present work, we opt to adopt a Gaussian gridding convolution function (GCF) that would mimic the Airy disk of 33$^{\prime\prime}$ and 51$^{\prime\prime}$ at 90 and 140\,$\mu$m, respectively. 
These parameters are chosen because they would reproduce the diffraction limit at the respective wavelengths in the resulting maps.
Our particular parameter choices are different from those made by \citet{ueta2017}, and therefore, would inevitably result in slightly different PSF shapes.
Hence, we need to derive our own surface brightness correction factors (see the next section; also see Appendix\,\ref{redo}).

We thus take advantage of the improved capability of FAST to optimize calibration and angular resolution of the resulting co-added maps. 
In the final map-making, we orient the resulting images to align with the scan direction (i.e., the image $y$-direction is parallel to the scan direction; also see Figure 3 of \cite{kawada2007}) and use the following options: Earth-shine/stray-light removal, 200\,s median filtering, 5-$\sigma$ clipping, and GCF corresponding to the Airy disk of 33$^{\prime\prime}$ and 51$^{\prime\prime}$ at 90 and 140\,$\mu$m.
The resulting FAST co-added maps are produced at the pixel scale of 8$^{\prime\prime}$\,pix$^{-1}$ (corresponding to roughly 1/4 to 1/6 of the angular resolution). 

\section{Surface Brightness Correction\label{S:corr}} % for the AKARI/FIS maps}

The absolute surface brightness calibration of the FIS instrument was done through (1) pre-launch laboratory measurements of a blackbody source which indicated a 5\,\% accuracy, and (2) in-orbit comparisons between measurements of IR cirrus regions without significant small-scale structures made by FIS and the DIRBE instrument on board the COBE satellite \citep{matsuura2011}.
Hence, the FIS data (both all-sky and slow-scan data) presently archived should give correct surface brightnesses of diffuse background emission. 
However, when aperture photometry was performed for a set of IR flux standard stars detected in the COBE/DIRBE-calibrated FIS slow-scan maps (processed by SS-Tool), the resulting fluxes came out to be roughly 40\,\% less than expected \citep{shirahata2009}.
This apparent flux underestimates were attributed to the slow transient response of the Ge:Ga detectors (e.g., \cite{kaneda2002}).

To alleviate this issue, \citet{shirahata2009} devised a method of flux correction for point sources detected in the AKARI/FIS slow-scan maps processed by SS-Tool. 
This point-source flux correction method is based on the the premise that the shape of the PSF is well-defined and invariant with the source flux.
This point-source flux correction method works because flux within an idealized infinite aperture can always be recovered by scaling the flux measured within a finite aperture (which corresponds to a part of the bright PSF core) by an appropriate aperture correction factor. 
However, such a correction method would not work in general for objects that are neither point-like nor diffuse (i.e., marginally extended, such as CDSs), because the surface brightness distribution of such an object is not known {\sl a priori} (and hence the aperture correction factor cannot be uniquely determined).

\citet{ueta2017} established a general procedure to correct directly the far-IR surface brightness distribution of AKARI/FIS slow-scan maps generated by SS-Tool or FAST.
This procedure is based on the empirical power-law FIS detector response function, $\mathcal{R}$, which is defined to be
\begin{equation}\label{fisresponse}
S_{ij, {\rm FIS}} = \mathcal{R}(S_{ij}) = c S_{ij}^n,
\end{equation}
where
$S_{ij, {\rm FIS}}$ is the measured surface brightness distribution in the archived, uncorrected FIS map,
$S_{ij}$ is the ``true'' surface brightness distribution of the mapped region of the sky in the far-IR,
$n$ and $c$ are, respectively, the power-law index and the scaling coefficient of the adopted power-law FIS response function, 
and 
$i$ and $j$ refer to the pixel position in the FIS map.

\begin{figure*}
\begin{center}
    \includegraphics[width=\textwidth]{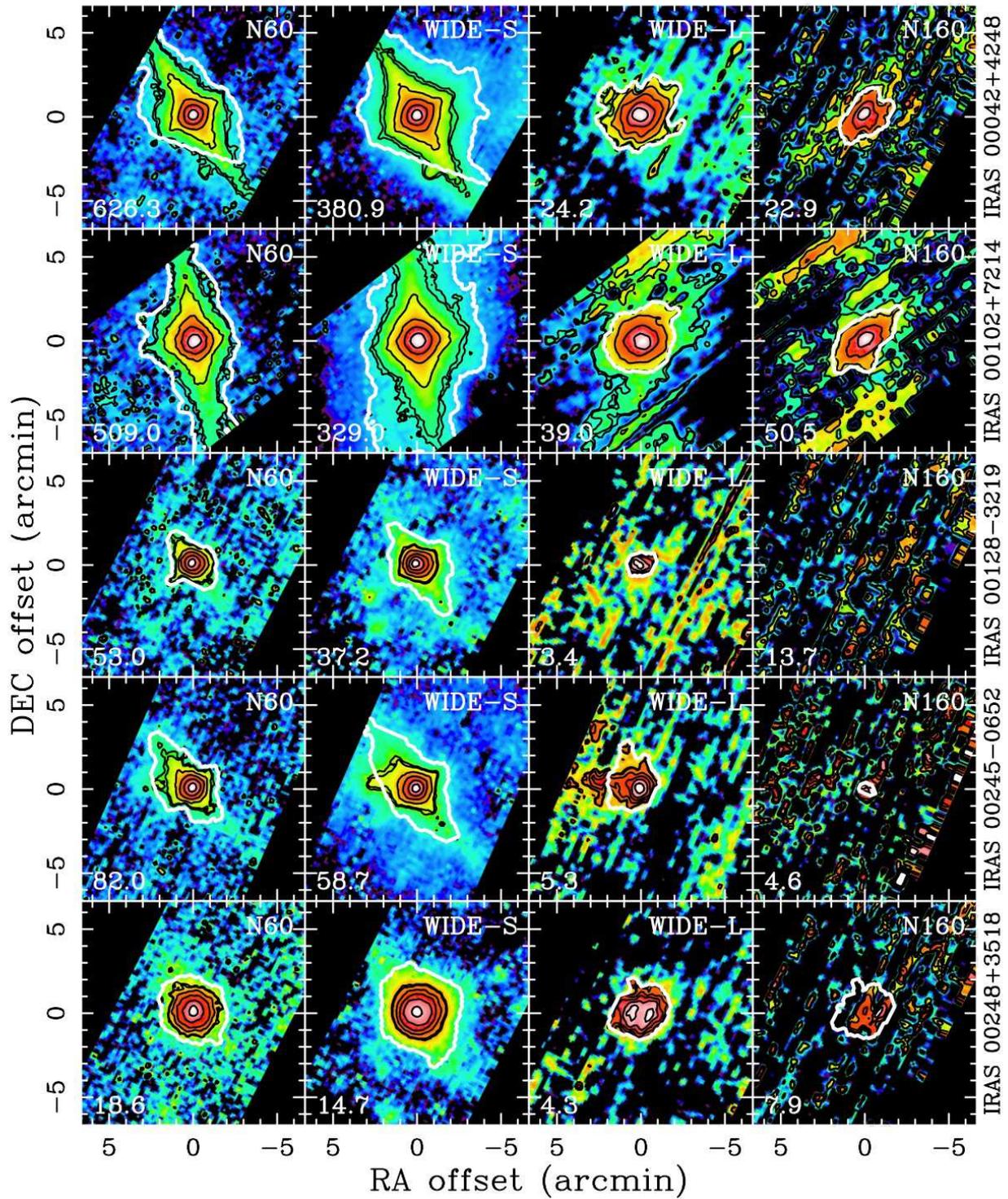}
\end{center}
\caption{%
Grayscale far-IR images of the MLHES target sources in the four AKARI/FIS bands (N60, WIDE-S, WIDE-L, and N160) from left to right, respectively.
Each panel shows a $13\farcm3 \times 13\farcm3$ region centered at the RA-DEC coordinates of the target, whose name is shown on the right margin.
White contours are 90, 70, 50, 30, 10, and 5\,\% of the peak surface brightness, which is given on the bottom left corner of each frame.
The black contour represents the adopted photometry boundary at each band.}\label{F:map1}
\end{figure*}

This FIS response function was determined to possess a power-law form given the observed scale-invariant characteristics of slow-scan maps (i.e., PSF shapes remain the same irrespective of the PSF brightness).
With this formulation, one can recover the true far-IR surface brightness distribution of the sky falling onto the FIS detectors via the inverse function, $\mathcal{R}^{-1}$, via
\begin{equation}\label{rescaling}
S_{ij} = \mathcal{R}^{-1}(S_{ij, {\rm FIS}}) = (S_{ij, {\rm FIS}}/c)^{1/n}.
\end{equation}
with the appropriately-determined $n$ and $c$ power-law scaling parameters.
The $n$ and $c$ parameters adopted in the present work are listed in Table\,\ref{nandcvalues}.
These parameters are slightly different from those previously determined by \citet{ueta2017}, because, as stated in the previous section, we opt to adopt specific GCF parameters to optimize the resulting PSF shape.
The derivation of the present $n$ and $c$ values is outlined in Appendix\,\ref{redo} in detail.
The resulting surface-brightness-corrected AKARI/FIS maps of the MLHES target sources are exhibited in Figure\,\ref{F:map1} and Supplementary Figures\,1 %\ref{F:map2} 
through 28 %\ref{F:map29}
\footnote{Supplementary Figures\,1 %\ref{F:map2} 
through 28 %\ref{F:map29} 
are available only in the on-line version.}.

\ifnum0=1 %%%%%%%%%%%%%%%%%%%%%%%%%%%%%%%%%%%%%%%%%%%%%%%%
\begin{figure}
\begin{center}
    \includegraphics[width=\columnwidth]{fig_panel02_bw.eps}
\end{center}
\caption{%
Grayscale far-IR images of the MLHES target sources in the four AKARI/FIS bands (N60, WIDE-S, WIDE-L, and N160) from left to right, respectively.
Image conventions are the same as Figure\,\ref{F:map1}.
}\label{F:map2}
\end{figure}

\begin{figure}[p]
\begin{center}
    \includegraphics[width=\columnwidth]{fig_panel03_bw.eps}
\end{center}
\caption{%
Grayscale far-IR images of the MLHES target sources in the four AKARI/FIS bands (N60, WIDE-S, WIDE-L, and N160) from left to right, respectively.
Image conventions are the same as Figure\,\ref{F:map1}.}\label{F:map3}
\end{figure}

\begin{figure}[p]
\begin{center}
    \includegraphics[width=\columnwidth]{fig_panel04_bw.eps}
\end{center}
\caption{%
Grayscale far-IR images of the MLHES target sources in the four AKARI/FIS bands (N60, WIDE-S, WIDE-L, and N160) from left to right, respectively.
Image conventions are the same as Figure\,\ref{F:map1}.
}\label{F:map4}
\end{figure}

\begin{figure}[p]
\begin{center}
    \includegraphics[width=\textwidth]{fig_panel05_bw.eps}
\end{center}
\caption{%
Grayscale far-IR images of the MLHES target sources in the four AKARI/FIS bands (N60, WIDE-S, WIDE-L, and N160) from left to right, respectively.
Image conventions are the same as Figure\,\ref{F:map1}.
}\label{F:map5} 
\end{figure}

\begin{figure}[p]
\begin{center}
    \includegraphics[width=\textwidth]{fig_panel06_bw.eps}
\end{center}
\caption{%
Grayscale far-IR images of the MLHES target sources in the four AKARI/FIS bands (N60, WIDE-S, WIDE-L, and N160) from left to right, respectively.
Image conventions are the same as Figure\,\ref{F:map1}.
}\label{F:map6}
\end{figure}

\begin{figure}[p]
\begin{center}
    \includegraphics[width=\textwidth]{fig_panel07_bw.eps}
\end{center}
\caption{%
Grayscale far-IR images of the MLHES target sources in the four AKARI/FIS bands (N60, WIDE-S, WIDE-L, and N160) from left to right, respectively.
Image conventions are the same as Figure\,\ref{F:map1}.
}\label{F:map7}
\end{figure}

\begin{figure}[p]
\begin{center}
    \includegraphics[width=\textwidth]{fig_panel08_bw.eps}
\end{center}
\caption{%
Grayscale far-IR images of the MLHES target sources in the four AKARI/FIS bands (N60, WIDE-S, WIDE-L, and N160) from left to right, respectively.
Image conventions are the same as Figure\,\ref{F:map1}.
}\label{F:map8}
\end{figure}

\begin{figure}[p]
\begin{center}
    \includegraphics[width=\textwidth]{fig_panel09_bw.eps}
\end{center}
\caption{%
Grayscale far-IR images of the MLHES target sources in the four AKARI/FIS bands (N60, WIDE-S, WIDE-L, and N160) from left to right, respectively.
Image conventions are the same as Figure\,\ref{F:map1}.
}\label{F:map9}
\end{figure}

\begin{figure}[p]
\begin{center}
    \includegraphics[width=\textwidth]{fig_panel10_bw.eps}
\end{center}
\caption{%
Grayscale far-IR images of the MLHES target sources in the four AKARI/FIS bands (N60, WIDE-S, WIDE-L, and N160) from left to right, respectively.
Image conventions are the same as Figure\,\ref{F:map1}.
}\label{F:map10}
\end{figure}

\begin{figure}[p]
\begin{center}
    \includegraphics[width=\textwidth]{fig_panel11_bw.eps}
\end{center}
\caption{%
Grayscale far-IR images of the MLHES target sources in the four AKARI/FIS bands (N60, WIDE-S, WIDE-L, and N160) from left to right, respectively.
Image conventions are the same as Figure\,\ref{F:map1}.
}\label{F:map11}
\end{figure}

\begin{figure}[p]
\begin{center}
    \includegraphics[width=\textwidth]{fig_panel12_bw.eps}
\end{center}
\caption{%
Grayscale far-IR images of the MLHES target sources in the four AKARI/FIS bands (N60, WIDE-S, WIDE-L, and N160) from left to right, respectively.
Image conventions are the same as Figure\,\ref{F:map1}.
}\label{F:map12}
\end{figure}

\begin{figure}[p]
\begin{center}
    \includegraphics[width=\textwidth]{fig_panel13_bw.eps}
\end{center}
\caption{%
Grayscale far-IR images of the MLHES target sources in the four AKARI/FIS bands (N60, WIDE-S, WIDE-L, and N160) from left to right, respectively.
Image conventions are the same as Figure\,\ref{F:map1}.
}\label{F:map13}
\end{figure}

\begin{figure}[p]
\begin{center}
    \includegraphics[width=\textwidth]{fig_panel14_bw.eps}
\end{center}
\caption{%
Grayscale far-IR images of the MLHES target sources in the four AKARI/FIS bands (N60, WIDE-S, WIDE-L, and N160) from left to right, respectively.
Image conventions are the same as Figure\,\ref{F:map1}.
}\label{F:map14}
\end{figure}

\begin{figure}[p]
\begin{center}
    \includegraphics[width=\textwidth]{fig_panel15_bw.eps}
\end{center}
\caption{%
Grayscale far-IR images of the MLHES target sources in the four AKARI/FIS bands (N60, WIDE-S, WIDE-L, and N160) from left to right, respectively.
Image conventions are the same as Figure\,\ref{F:map1}.
}\label{F:map15}
\end{figure}

\begin{figure}[p]
\begin{center}
    \includegraphics[width=\textwidth]{fig_panel16_bw.eps}
\end{center}
\caption{%
Grayscale far-IR images of the MLHES target sources in the four AKARI/FIS bands (N60, WIDE-S, WIDE-L, and N160) from left to right, respectively.
Image conventions are the same as Figure\,\ref{F:map1}.
}\label{F:map16}
\end{figure}

\begin{figure}[p]
\begin{center}
    \includegraphics[width=\textwidth]{fig_panel17_bw.eps}
\end{center}
\caption{%
Grayscale far-IR images of the MLHES target sources in the four AKARI/FIS bands (N60, WIDE-S, WIDE-L, and N160) from left to right, respectively.
Image conventions are the same as Figure\,\ref{F:map1}.
}\label{F:map17}
\end{figure}

\begin{figure}[p]
\begin{center}
    \includegraphics[width=\textwidth]{fig_panel18_bw.eps}
\end{center}
\caption{%
Grayscale far-IR images of the MLHES target sources in the four AKARI/FIS bands (N60, WIDE-S, WIDE-L, and N160) from left to right, respectively.
Image conventions are the same as Figure\,\ref{F:map1}.
}\label{F:map18}
\end{figure}

\begin{figure}[p]
\begin{center}
    \includegraphics[width=\textwidth]{fig_panel19_bw.eps}
\end{center}
\caption{%
Grayscale far-IR images of the MLHES target sources in the four AKARI/FIS bands (N60, WIDE-S, WIDE-L, and N160) from left to right, respectively.
Image conventions are the same as Figure\,\ref{F:map1}.
}\label{F:map19}
\end{figure}

\begin{figure}[p]
\begin{center}
    \includegraphics[width=\textwidth]{fig_panel20_bw.eps}
\end{center}
\caption{%
Grayscale far-IR images of the MLHES target sources in the four AKARI/FIS bands (N60, WIDE-S, WIDE-L, and N160) from left to right, respectively.
Image conventions are the same as Figure\,\ref{F:map1}.
}\label{F:map20}
\end{figure}

\begin{figure}[p]
\begin{center}
    \includegraphics[width=\textwidth]{fig_panel21_bw.eps}
\end{center}
\caption{%
Grayscale far-IR images of the MLHES target sources in the four AKARI/FIS bands (N60, WIDE-S, WIDE-L, and N160) from left to right, respectively.
Image conventions are the same as Figure\,\ref{F:map1}.
}\label{F:map21}
\end{figure}

\begin{figure}[p]
\begin{center}
    \includegraphics[width=\textwidth]{fig_panel22_bw.eps}
\end{center}
\caption{%
Grayscale far-IR images of the MLHES target sources in the four AKARI/FIS bands (N60, WIDE-S, WIDE-L, and N160) from left to right, respectively.
Image conventions are the same as Figure\,\ref{F:map1}.
}\label{F:map22}
\end{figure}

\begin{figure}[p]
\begin{center}
    \includegraphics[width=\textwidth]{fig_panel23_bw.eps}
\end{center}
\caption{%
Grayscale far-IR images of the MLHES target sources in the four AKARI/FIS bands (N60, WIDE-S, WIDE-L, and N160) from left to right, respectively.
Image conventions are the same as Figure\,\ref{F:map1}.
}\label{F:map23}
\end{figure}

\begin{figure}[p]
\begin{center}
    \includegraphics[width=\textwidth]{fig_panel24_bw.eps}
\end{center}
\caption{%
Grayscale far-IR images of the MLHES target sources in the four AKARI/FIS bands (N60, WIDE-S, WIDE-L, and N160) from left to right, respectively.
Image conventions are the same as Figure\,\ref{F:map1}.
}\label{F:map24}
\end{figure}

\begin{figure}[p]
\begin{center}
    \includegraphics[width=\textwidth]{fig_panel25_bw.eps}
\end{center}
\caption{%
Grayscale far-IR images of the MLHES target sources in the four AKARI/FIS bands (N60, WIDE-S, WIDE-L, and N160) from left to right, respectively.
Image conventions are the same as Figure\,\ref{F:map1}.
}\label{F:map25}
\end{figure}

\begin{figure}[p]
\begin{center}
    \includegraphics[width=\textwidth]{fig_panel26_bw.eps}
\end{center}
\caption{%
Grayscale far-IR images of the MLHES target sources in the four AKARI/FIS bands (N60, WIDE-S, WIDE-L, and N160) from left to right, respectively.
Image conventions are the same as Figure\,\ref{F:map1}.
}\label{F:map26}
\end{figure}

\begin{figure}[p]
\begin{center}
    \includegraphics[width=\textwidth]{fig_panel27_bw.eps}
\end{center}
\caption{%
Grayscale far-IR images of the MLHES target sources in the four AKARI/FIS bands (N60, WIDE-S, WIDE-L, and N160) from left to right, respectively.
Image conventions are the same as Figure\,\ref{F:map1}.
}\label{F:map27}
\end{figure}

\begin{figure}[p]
\begin{center}
    \includegraphics[width=\textwidth]{fig_panel28_bw.eps}
\end{center}
\caption{%
Grayscale far-IR images of the MLHES target sources in the four AKARI/FIS bands (N60, WIDE-S, WIDE-L, and N160) from left to right, respectively.
Image conventions are the same as Figure\,\ref{F:map1}.
}\label{F:map28}
\end{figure}

\begin{figure}[p]
\begin{center}
    \includegraphics[width=\textwidth]{fig_panel29_bw.eps}
\end{center}
\caption{%
Grayscale far-IR images of the MLHES target sources in the four AKARI/FIS bands (N60, WIDE-S, WIDE-L, and N160) from left to right, respectively.
Image conventions are the same as Figure\,\ref{F:map1}.
}\label{F:map29}
\end{figure}
\fi %%%%%%%%%%%%%%%%%%%%%%%%%%%%%%%%%%%%%%%%%%%%%%%%%%%

\section{Photometry}

After AKARI/FIS maps of the MLHES target sources are processed and their surface brightnesses are corrected as described in the previous sections, we measure the total flux density for each target source in each of the four AKARI/FIS band using the following three-step procedure, scripted with Python aided by the Astropy package.

{\bf Step 1:}
We determine the first-pass sky surface brightness, $I_{\rm sky}$, and its uncertainty, $\sigma_{\rm sky}$, by taking the median and standard deviation of the entire FIS map, respectively, using 3-$\sigma$ clipping.
We then define the extent of the target source by adopting a region-growth algorithm.
With this algorithm, a given region starting with the seed pixel at the target coordinates is grown by finding and incorporating all pixels that are connected neighbors to the region pixels as long as their pixel surface brightness value registers greater than the tentative sky value by 2\,$\sigma_{\rm sky}$ or more 
(i.e., $\ge (I_{\rm sky} + 2\,\sigma_{\rm sky}$)).

{\bf Step 2:} 
We then establish the sky regions adjacent to the grown target region in the surface brightness map (i.e., fore and aft directions along the scan path) to determine the second-pass $I_{\rm sky}$ and $\sigma_{\rm sky}$ by taking the median and standard deviation of the pixels within the sky regions using 3-$\sigma$ clipping.
With the re-determined $I_{\rm sky}$ and $\sigma_{\rm sky}$, we use the region-growth algorithm again to re-define the extent of the target source.
The source region is grown to include all connected pixels registering greater than the updated sky value by 2\,$\sigma_{\rm sky}$ or more.
The adopted 2\,$\sigma_{\rm sky}$ threshold is chosen from comparisons among 1, 1.5, 2, 2.5, and 3\,$\sigma_{\rm sky}$ to find the optimum level that would recover the faintest surface brightness of the target sources while not including spurious emission from the sky or nearby sources.

This iteration is intended to make sure that the extent of the target source comes as low in the surface brightness as possible and that the sky regions do not overlap with the derived extent of target source.
For this region-grow method to work, however, the target sources are assumed to have a single emission peak at the target position with their surface brightness distributions in general radially decreasing monotonically.
This assumption, however, is not always true because the surface brightness distribution can vary considerably when the circumstellar shell possesses noticeable structures.
Thus, we also visually inspect the results of the second-pass region-grow algorithm to check especially if real structures are missed or spurious structures are included.
There are indeed cases where 
(1) what appears to be a nearby source is present in the immediate vicinity of the target (e.g., I00042\footnote{From here, we refer to our targets by the abbreviated IRAS designation, 
an ``I" followed by the first 5 digits of the IRAS designation, or alternative name when there is no IRAS designation.}, 
I00245, I04020, I16011, I21197, I21440, and I27173), and
(2) the target object is confused by the strong large-scale background emission (e.g., I3062 and I21419).
These cases are handled on an individual basis by manually modifying the photometry boundary to exclude confused parts.
In general, nearby sources are deemed real when they appear in more than one consecutive bands and/or exhibit different emission characteristics than the target source itself (i.e., the way the brightness varies across bands is dissimilar).

Also, there are cases where the background is too noisy to define a reasonable boundary at the adopted 2\,$\sigma_{\rm sky}$ threshold.
In such cases, the target sources may not be automatically recognized even though they may appear obvious to human eyes.
If that is the case, we adopt the boundary determined at the waveband of the best data quality (typically the waveband of the shortest wavelength where the boundary was algorithmically determined; i.e., the WIDE-L boundary is adopted for the N160 band, and the WIDE-S boundary is adopted for the WIDE-L and N160 bands).
At any rate, whenever we adopt the photometry boundary that is not algorithmically determined (i.e., there is some manual intervention), it is noted in Table\,\ref{tab:fluxes}.
These manual modifications of the photometry boundary typicaly introduce a difference of at most a few\,\% (nearby source cases) to a few tens of \% (background confusion cases).

{\bf Step 3:} 
We then compute the flux density of the target source and its uncertainty using the following formulae:
\begin{eqnarray}
F_{\nu}
&=&
\sum_{i,j}^{{\rm obj}} \mbox{img}(i,j)
-
\frac{N_{\rm obj}}{N_{\rm sky}}
\sum_{i,j}^{{\rm sky}} \mbox{img}(i,j)
\label{eq3}
\\
\sigma_{F_{\nu}}^2
&=&
\sum_{i,j}^{{\rm obj}} \mbox{err}(i,j)^2
-
\frac{N_{\rm obj}}{N_{\rm sky}}
\sum_{i,j}^{{\rm sky}} \mbox{err}(i,j)^2
\nonumber\\
&~&
~~~~~~~~~~~~
+
N_{\rm obj} \sigma_{\rm sky}^2
+
\frac{N_{\rm obj}^2}{N_{\rm sky}} \sigma_{\rm sky}^2
\label{eq4}
\end{eqnarray}
where img$(i,j)$ is the re-calibrated surface brightness map and err$(i,j)$ is the associated uncertainty map, 
$N_{\rm obj}$ and $N_{\rm sky}$ are the number of pixels within the extent of the target object (``obj") and of the corresponding sky region (``sky"), respectively,
and
$\sigma_{\rm sky}$ is the standard deviation in the sky regions.
The second-pass sky regions immediately before and after the second-pass grown target source region are referred to as the sky regions (``sky"), while the second-pass grown target source region itself is referred to as the target object region (``obj").
Similarly, the corresponding regions in the associated error map are referred to as in the same way (as the ``sky" and ``obj"), respectively.
The err$(i,j)$ term corresponds solely to the uncertainties of the detector signal per sampling, while $\sigma_{\rm sky}$ is the fluctuation of the observed sky emission (i.e., confusion noise and uncertainties of the detector signal per sampling):
the $\sigma_{\rm sky}$ term is often a significant component in the far-IR.

The second term of eqn (\ref{eq3}) subtracts the sky contribution in the target area.
The first term in eqn (\ref{eq4}) is a mixture of 
the photon shot noise due to signals from the target,
photon shot noise due to sky emission,
readout noise, 
dark current noise,
and 
uncertainties in flatfielding.
The third term in eqn (\ref{eq4}) refers to the uncertainty that arises in the target region due to sky fluctuation (i.e., 
the photon shot noise due to sky emission,
confusion noise,
readout noise, 
dark current noise,
and 
uncertainties in flatfielding,
assuming that similar sky fluctuation continues into the target region).
Finally, the fourth term in eqn (\ref{eq4}) is the uncertainty due to sky subtraction.
Thus, we are actually double-counting
the photon shot noise due to sky emission,
readout noise, 
dark current noise,
and 
uncertainties in flatfielding.
Hence, the second term in eqn (\ref{eq4}) is included to compensate for those doubly-added noise terms.
In addition,
surface brightnesses of the re-calibrated map and its associated uncertainty map are converted from the native MJy\,sr$^{-1}$ to Jy\,pix$^{-1}$ units given the adopted pixel scale at 8$^{\prime\prime}$\,pix$^{-1}$ to yield flux densities in the end.

The measured flux densities of the MLHES target sources in the four AKARI far-IR bands are presented in Table\,\ref{tab:fluxes}.
The listed flux densities are expressed in scientific notation,
${\rm a} \pm {\rm b} ({\rm c})$, 
where 
${\rm a} \pm {\rm b}$ is the absolute value and 
${\rm c}$ is the power index of base ten, i.e., $(a \pm b) \times 10^{c}$.
Here, readers are reminded that these flux densities are the sum of the flux densities from the central star and the CDS.
In this list, we also note if the photometry aperture boundary set by the procedure outlined above is manually adjusted (e.g., a nearby object needs to be removed) or adopted from a shorter waveband (e.g., the sky is too noisy to define a decent aperture in the present band).

In addition, we also present the far-IR sky surface brightnesses and their uncertainties at the positions of the MLHES target sources (Table\,\ref{tab:skySB}).
These sky values are computed by taking the median and standard deviation of the sky-only maps,
which are created by taking the difference between the surface-brightness-corrected FIS maps with the median-filter on (i.e., the sky emission is subtracted from the map) and off (i.e., the sky emission remains in the map).

\section{Discussion}

\subsection{AKARI/FIS Images of Evolved Star CDSs}

One of the main objectives of the present investigation is to understand the history of AGB mass loss via the far-IR surface brightness distribution of the cold dust component in the evolved star CDSs.
This is because the CDSs are the direct consequences of mass loss.
Upon quickly inspecting the resulting AKARI/FIS images (Figure\,\,\ref{F:map1} and Supplementary Figures\,\ref{F:map2} through \ref{F:map29}), we immediately notice that the central star is in general still the dominant emission source even at these far-IR wavelengths.
This is especially true in the SW bands, and quite often so even in the WIDE-L band.
Nevertheless, we find many sources with a faint extended CDS.

While we leave detailed analyses into the CDS structures to the next installment of the present series, we briefly comment on the general characteristics of the CDS structures that we can observe from a quick inspection.
Many sources appear to be associated with a round extended CDS, some are very obvious (up to about $5^{\prime}$ radius) and others are less so.
Such sources, of which a few have been previously presented elsewhere, include
I04330, 
I05524 \citep{ueta2008}, 
I10329 \citep{izumiura2011}, 
I10350,
I10580,
I12427,
I13462,
I15465,
I15094,
I18537, and
I23558 \citep{ueta2010}.

There are many sources whose emission core appears to be more extended than that of the PSF at varying degrees.
These objects can also be identified as those whose internal contours are more widely spaced.
Such sources include
I01037,
I02108,
I02522,
I03062,
I03374,
I04361,
I05251,
I06176,
I09452,
I10223,
I10416,
I11385,
I12380,
I13370
I14003,
I14219,
I16011,
I17028,
I19390,
I19434,
AFGL2688,
NGC7027,
I21419,
I22035,
I22196, and
I23166.
Those showing an especially extended emission core are
I00248,
I11119,
I18517,
I19574, and
I20120.
Some of these sources have been found extended in resolved far-IR images taken by Herschel (e.g., 
I00248, I10329, and I19390;
\cite{kerschbaum2010}).
Many others seem at least marginally extended, and they need more detailed analyses
before concluding anything.

Another group of sources that is worth mentioning here is those that show some indications of interactions between the CDS and ISM.
Such cases have been detected in the far-IR around evolved stars and suggested to exist rather commonplace (e.g., \cite{ueta2006,ueta2010,ueta2011,jorissen2011,cox2012,mayer2013,mayer2014}).
Such objects are
I03463,
I04459,
I13001,
I16255.
I18216,
I20038,
I21412,
I22017,
I22272, and
I23416.

GK Per is an unusual magnetic cataclysmic variable star that went into a nova in 1901 and the remnant of the outburst is presently seen as a nebula of about 100$^{\prime\prime}$ diameter \citep{slavin95}.
This system, recently undergoing outbursts roughly every 3 years, exhibited a longer-than-normal outburst in 2006 \citep{evans06}.
This motivated us to observe this system in the far-IR to see if there is any cold thermal dust emission.
However, no appreciable far-IR emission was detected in all four AKARI bands.

Overall, we typically find at least a couple of marginally extended sources in each of the 29 panels of images presented.
Thus, about 60-70\,\% of the observed objects, roughly speaking, appear to exhibit some kind of extension of the CDS in the far-IR.
Many extended objects, especially those listed above, appear to be large enough that its inner shell structure is resolved and can be revealed after the effects of the bright central star is suppressed.
Hence, careful central star removal will have to be performed to fully assess the nature of the extended CDSs of evolved stars in the far-IR.
Such an analysis will be the main topic of the next installment of the present series of papers (Torres et al.\, {\sl in preparation}).

\subsection{Comparison with the IRAS PSC and AKARI/FIS BSC Entries}

One of the main purposes of AKARI is to perform a census of far-IR objects in the whole sky \citep{Yamamura_2009}.
Hence, there must be corresponding catalog entries for all MLHES sources, as the MLHES targets are essentially selected from the IRAS catalog (i.e., bright enough for AKARI).
Thus, as a quick consistency check,
we compare the flux densities of the MLHES targets determined by our method
with their counterparts in the AKARI/FIS Bright Source Catalog Ver.\,2 (BSCv2; \cite{Yamamura_2016})\footnote{Available via http://www.ir.isas.jaxa.jp/AKARI/Archive/.}.
We compare only those whose quality flag is 3 (the presence of the source is confirmed and its flux determined to be valid in BSCv2).

Here, we need to remind ourselves that the flux density entries in BSCv2 are 
measured by yet another independent method, dedicated to produce BSC flux entries
with the assumption that the signal is always caused by a point source \citep{Yamamura_2009}.
Moreover, BSCv2 flux entries are calibrated by a direct comparison 
with the point-source calibration standard fluxes 
and not from the surface brightness maps as in our method presented in this work .
These differences practically mean that the BSCv2 flux density entries for the MLHES target sources tend to miss contributions from the extended parts of the CDS around the central star. 
However, the severities of this potential flux density underestimate would depend on the actual appearance of the target sources in the AKARI focal plane.
Hence, BSCv2 entries may not always underestimate flux densities of the MLHES sources.

\begin{figure}
\begin{center}
    \includegraphics[width=\columnwidth]{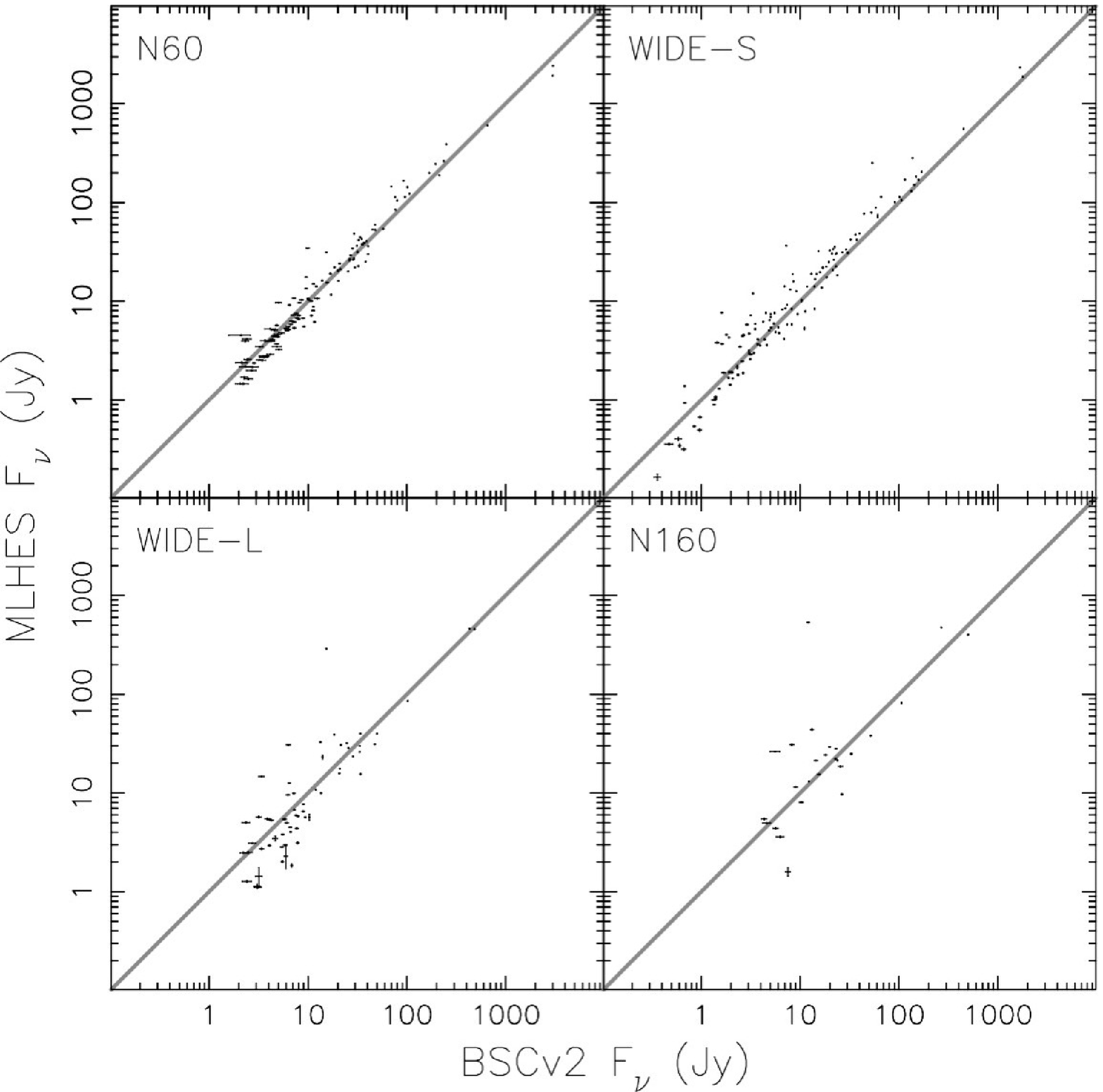}
\end{center}
\caption{%
Comparisons between our photometry measurements and BSCv2 flux density catalog entries in each of the four AKARI bands, 
N60, WIDE-S, WIDE-L, and N160, from top left to bottom right, in a log-log plot.
Uncertainties are shown as error bars.
The gray line shows a locus of points at which both flux density values are equal (i.e., the $y=x$ line).
}\label{F:photcompbsc}
\end{figure}

Figure \ref{F:photcompbsc} shows the direct comparison between the photometry results of the MLHES targets and their counterparts in BSCv2.
In general, our photometry results are consistent with their BSCv2 counterparts, as seen from the distribution of data points that follows the $y=x$ line in gray. 
However, there is also a relatively large scatter in the distribution of the data points.
The width of the scatter is therefore interpreted as the manifestation of this intrinsic difference between the way photometry is done by us and for BSCv2.
While there is this subtle difference between our measurements and BSC entries, 
blind power-law fitting of all the data points yield
$F_{\rm N60(MLHES)}    = 1.311\pm0.005 F_{\rm N60(BSC)}^{(0.958\pm0.001)}$,
$F_{\rm WIDE-S(MLHES)} = 1.116\pm0.003 F_{\rm WIDE-S(BSC)}^{(1.063\pm0.001)}$,
$F_{\rm WIDE-L(MLHES)} = 0.759\pm0.004 F_{\rm WIDE-L(BSC)}^{(1.063\pm0.001)}$, and
$F_{\rm N160(MLHES)}   = 0.802\pm0.009 F_{\rm N160(BSC)}^{(1.069\pm0.002)}$,
for N60, WIDE-S, WIDE-L, and N160, respectively.
The derived power-law indices are indeed fairly close to unity, and hence, we consider that our photometry results are reasonable.

\subsection{Far-IR Colors of the Evolved Star CDS}

After the IRAS all-sky survey, \citet{vh88} proposed that IRAS two-color diagrams can be used to distinguish C- and O-rich CDSs and put these stars align with the evolutionary sequence.
Here, we attempt similar exercise with AKARI two-color diagrams.
However, no colors based purely on AKARI flux densities yields any informative distribution of source in any AKARI two-color space, because
(1) AKARI's bands are defined over a relatively smaller range of wavelengths and 
(2) there is a significant overlap between the N60 and WIDE-S bands and the WIDE-L and N160 bands, respectively.
Hence, after experimenting with various photometry data in the infrared, we settle with the WISE W2 band \citep{wise}, because WISE data are among the latest and this band appears least likely to be contaminated by some line emission.

\begin{figure}
\begin{center}
    \includegraphics[width=\columnwidth]{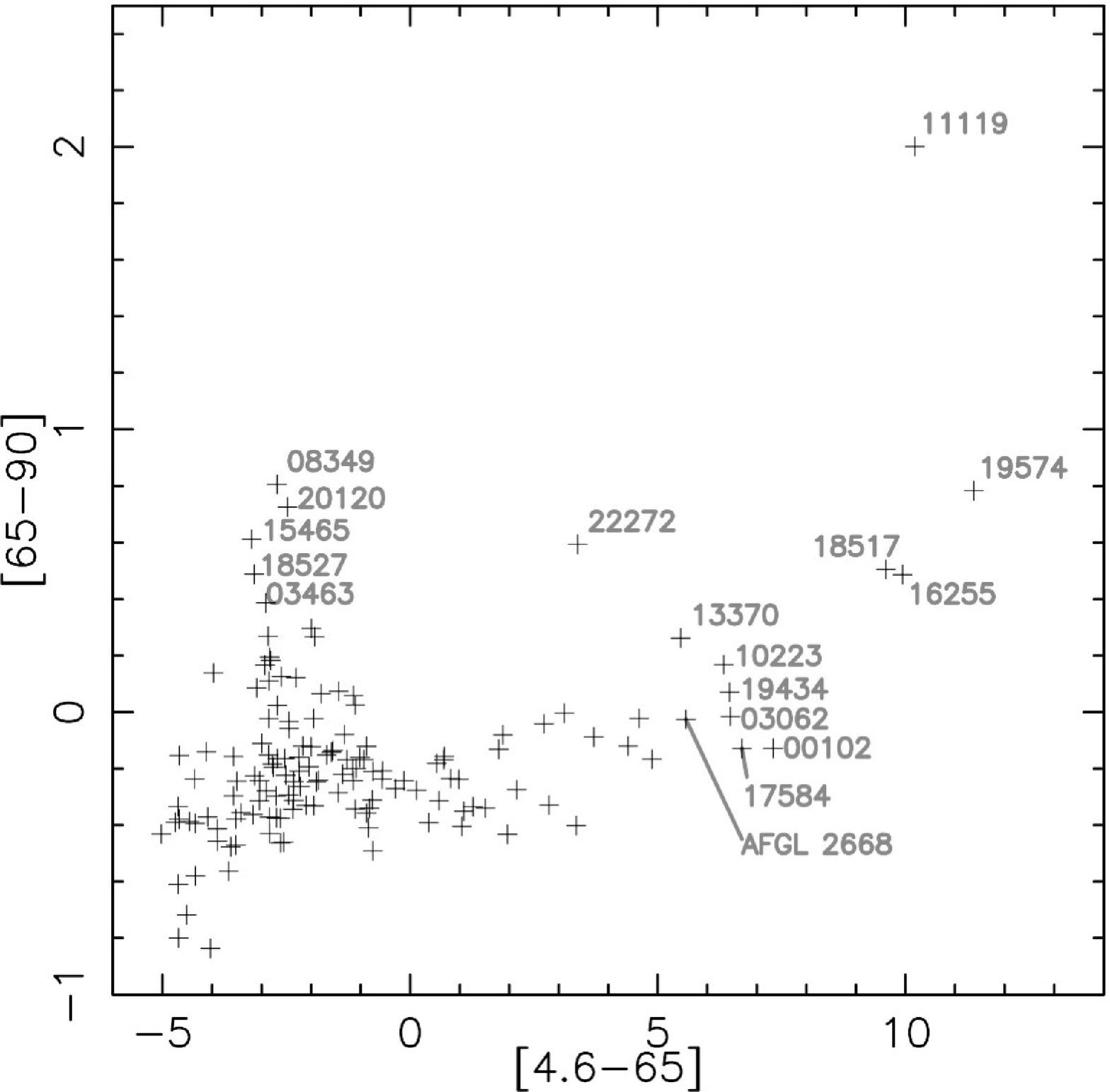}
\end{center}
\caption{%
The distribution of the MLHES targets in the [65-90] vs.\ [4.6-65] two-color diagram.
Sources found along the tips of horizontal and vertical loci emanating from the clustering of sources tend to be associated with an extended CDS and their names are shown.
}\label{F:2color}
\end{figure}

Figure\,\ref{F:2color} shows the distribution of the MLHES sources in the [65-90] vs.\ [4.6-65] two-color diagram\footnote{The color is defined to be $[{\rm a}-{\rm b}]=-2.5 \log \left(F_{\rm a}/F_{\rm b}\right)$, where $F_{\rm a}$ and $F_{\rm b}$ are flux densities in the band a and b, respectively.}.
Many sources cluster in the region of [65-90]$< 0$ and [4.6-65]$< 0$.
From this clustering of sources in the [65-90]-[4.9-65] color space, it appears that there are two loci, one horizontal and one vertical, along which other sources are found.
In Figure\,\ref{F:2color}, sources located along the tips of these loci are identified.
Interestingly, sources at the tip of the horizontal locus (i.e., those with redder [4.6-65] colors) are found to be exclusively PNe, while those at the tip of the vertical locus (i.e., those with redder [65-90] colors) are AGB stars already associated with extended CDSs.

The [65-90] color essentially determines where the dust emission peak is in the spectral energy distribution.
That is, if the dust emission peaks at the blueward of 65\,$\mu$m the color is blue (negative), at around $65-90$\,$\mu$m the color is gray (around zero), and the redward of 90\,$\mu$m the color is red (positive).
Meanwhile, the [4.6-65] color compares the relative strength between the stellar emission and the dust emission.
While the CDS is being developed during the mass-losing phase the [4.6-65] color is still quite blue, because the flux at 4.6\,$\mu$m from the cold central star is strong.
However, when the CDS is fully developed the [4.6-65] color would be reddened, because the flux at 90\,$\mu$m from the CDS is now full-fledged.
Hence, this two-color diagram allows us to determine which sources are likely to have an extended CDS and perform central source subtraction in characterizing further the intrinsic CDS structures for the second installment of the series.

\subsection{Newly Resolved Sources}

%The size of the primary mirror of AKARI is smaller than that of Herschel (0.685\,m as opposed to 3.4\,m).
%Thus, AKARI's spatial resolution is relatively marginal ($30^{\prime\prime} - 50^{\prime\prime}$ at $90 - 140$\,$\mu$m) in comparison with Herschel's.
%However, AKARI's sensitivity is comparable to or better than Herschel's, especially for diffuse emission, because AKARI's 
%primary mirror was cryogenically cooled \citep{akari}.
%However, 
AKARI's spatial resolutions are superior with respect to those previously achieved by IRAS and ISO.
Indeed, we find 29 instances in 28 targets in which another source is resolved in the immediate vicinity of the MLHES target sources (within the $13\farcm3 \times 13\farcm3$ FoV shown; two nearby sources are found in one case).
These newly-resolved nearby sources are most likely blended with the target sources themselves when previously observed with IRAS and/or ISO.
Most of these newly resolved sources are point sources and fainter than the main target source.
However, in a few cases the newly found sources are almost comparable to the main target in brightness.
Furthermore, there are several cases in which these nearby sources appear extended (sometimes more than the main target).

Such newly resolved sources are summarized in Table\,\ref{tab:extrasources}, with their coordinates, photometry results, and their nature.
We use 
the Simbad database\footnote{http://simbad.u-strasbg.fr/simbad/}
to query the AKARI coordinates of these newly resolved sources
to see if any of these sources have been recognized previously.
When a previously known source is found within the FWHM of the PSF, we associate the AKARI detection to such a source.
Out of 29 objects newly resolved by AKARI, 17 are found to have been previously known as identified in Table\,\ref{tab:extrasources}.
Of which, 14 are identified to be galaxies, while the rest is simply known as an IR or a radio source.
%(I05367 and I18525) %RAS 05367$-$8627 and IRAS 18525$+$3643) 
%or a radio source %(NVSS J072806$+$455415).
Below, we briefly comment on each of these newly resolved sources.

Those associated with 
I00042 ($3^{\prime}$ to S),
I00245 ($2^{\prime}$ to E),
I04020 ($3^{\prime}$ to SW, extended),
I04387 ($4^{\prime}$ to NE),
I05096 ($2^{\prime}$ to S),
I07245 ($5^{\prime}$ to S, the farther of the two),
I10323 ($6^{\prime}$ to S),
I14567 ($4\farcm5$ to W),
I18527 ($2^{\prime}$ to S, the closer of the two)
I21197 ($2^{\prime}$ to NE),
I21440 ($3^{\prime}$ to NE, extended), and
I23013 ($4\farcm5$ to N)
are previously ``unknown" (i.e., no corresponding object found after Simbad queries).
Given that the majority of the previously known sources are galaxies, these presently unknown sources are most probably galaxies.
However, two of these presently unknown sources are extended: we will investigate their nature in future.

\section{Summary}

Using the FIS on-board the AKARI IR astronomical satellite, we carried out a far-IR imaging survey of the CDSs of 144 evolved stars as one of the AKARI mission programmes that supplemented the satellite's all-sky survey observations.
With this survey we collected far-IR images of roughly $10^{\prime} \times 40^{\prime}$ or $10^{\prime} \times 20^{\prime}$ around the target evolved stars at 65, 90, 140, and 160\,$\micron$ (Figure\,\,\ref{F:map1} and Supplementary Figures\,\ref{F:map2} through \ref{F:map29}). 

We detect far-IR emission from all but one object (nova GK Per) at the spatial resolution about $30^{\prime\prime} - 50^{\prime\prime}$. 
Roughly $60 - 70$\,\% of the detected sources are found extended, to be followed up with more rigorous central source subtraction to isolate the CDS,
which will be discussed in the second installment of the present series.
With AKARI's better spatial resolution in comparison with the previous IRAS and ISO images, 29 new sources are now resolved separately around 28 target sources in the very vicinity of the targets.
About half of the newly-resolved sources (17/29) turn out to be previously known sources, and the majority (14/17) happens to be galaxies.
Hence, the majority of previously unknown sources is also expected to be galaxies. 
However, there are two extended nearby sources that are previously unidentified, and they will be followed up in future.

The results of photometry measurements (Table\,\ref{tab:fluxes}) are reasonable with respect to the entries in the AKARI/FIS BSCv2, despite the fact that the targets are assumed to be point-sources when the BSCv2 catalogue flux densities were computed.
We also tabulate the measured far-IR sky surface brightness in MJy\,sr$^{-1}$ at the positions of the MLHES targets 
as a future reference (Table\,\ref{tab:skySB}).
The AKARI-WISE two-color diagram of [65-90] vs.\ [4.6-65]
%augmented by the WISE W2 band flux densities at 4.6\,$\mu$m 
appears to indicate the progression of the evolution of the CDSs as the CDS becomes more extended and colder (i.e., redder) as time passes.
Hence, this two-color diagram can be an ideal tool to determine which sources are likely to have an extended CDS and perform PSF subtraction in characterizing further the intrinsic CDS structures for the second installment of the series.

\begin{ack}
This research is based on observations with AKARI, 
a JAXA project with the participation of ESA.
TU recognizes partial support from the Japan Society 
for the Promotion of Science (JSPS) through a FY2013 
long-term invitation fellowship program (L13518). 
RLT was partially supported by the NSF EAPSI Program 
(OISE-1209948).
This research has made use of the SIMBAD database, 
operated at CDS, Strasbourg, France,
as well as data products from the Wide-field IR Survey Explorer, 
which is a joint project of the University of California, Los Angeles, and the Jet Propulsion Laboratory/California Institute of Technology, funded by NASA.
Authors also thank the anonymous referee for his/her valuable comments to improve on the clarity of the manuscript.
\end{ack}

\section*{Supplementary Data} 

The following supplementary data is available at PASJ online. 

Supplementary Figures 1--28.

\appendix 
\section{Revised Scaling Parameters for the Power-Law FIS Response Function\label{redo}}

\subsection{Background}

The AKARI/FIS slow-scan maps need to be re-calibrated depending on the science to be done with the maps unless the science targets are the diffuse background emission, with which the maps are absolutely calibrated \citep{doi2015}.
For point sources, measured fluxes need to be re-scaled to yield corrected fluxes \citep{shirahata2009}, whereas for slightly extended objects such as circumstellar shells, surface brightness maps themselves must be re-scaled before fluxes can be correctly measured \citep{ueta2017}.
These re-calibration steps are necessary because the sensitivity of the FIS detector arrays is dependent on the strength of the incoming signal \citep{matsuura2011}.

The basis of the surface brightness correction method for AKARI/FIS maps is that the PSF shape always remains the same irrespective of the source flux (i.e., scale invariance). 
This particular characteristic of the FIS detector arrays permit us to describe the FIS response function as a power-law.
Therefore, a specific set of parameters for the adopted FIS power-law function (the $n$ and $c$ values; \S\,\ref{S:corr}) would specifically describe the way the FIS detector arrays responded to incoming signals to yield the specific PSF shape \citep{ueta2017}.

In characterizing this FIS response function, \citet{ueta2017} previously used the Gaussian gridding convolution function (GCF) of the beam (FWHM) size of $30^{\prime\prime}$ and $50^{\prime\prime}$ for the SW and LW bands, respectively,
upon converting the archived AKARI/FIS TSD into 2-D co-added maps with FAST.
These numbers were chosen to keep the consistency between the AKARI FIS All-Sky Survey (AFASS) images and the FIS slow-scan maps.
However, this particular parameter choice was not the best 
if the angular resolution of the resulting co-added images of the FIS slow-scan maps was to be optimized. 
This is because the kernel size for the point-spread-function (PSF) profile adopted was roughly 1.7 times greater than the intrinsic diffraction limit of AKARI at the wavelength of the wide bands at 90 and 140\,$\mu$m.

Thus, we need to re-define parameters of the FIS power-law scaling function that are optimized for the present investigation in which the spatial resolution is the utmost importance.
In other words, in the present investigation we aim to recover the intrinsic diffraction limit of AKARI at the wavelength of the wide bands at 90 and 140\,$\mu$m, and hence, we have to use the GCF that represents to the PSF shape at AKARI's intrinsic diffraction limit.
Hence, the $n$ and $c$ parameters of the FIS power-law scaling function (s; \S\,\ref{S:corr}) must be updated according to the newly adopted PSF shape at the intrinsic diffraction limit.
%, following the same procedure outlined by \citet{ueta2017}.

\subsection{Super-PSF of the FIS Slow-Scan Maps at AKARI's Diffraction Limit}

For the present analysis in which we aim to produce AKARI/FIS images of the highest spatial resolution, we adopt the GCF mimicking the Airy disk of 33$^{\prime\prime}$ and 51$^{\prime\prime}$ for SW and LW, respectively (i.e., the intrinsic diffraction limit of AKARI at these wavelengths).
The effective Airy disk size (defined to be of the 1/$e^2$ diameter) measured from the resulting super-PSFs turns out to be
49\farcs1
and
76\farcs7,
while the effective FWHM is
23\farcs3
and
34\farcs9.
Thus, the effective FWHM size is roughly a factor of 2 smaller in the present study with respect to those used by \citet{ueta2017}.

\begin{figure}
\begin{center}
    \includegraphics[width=\columnwidth]{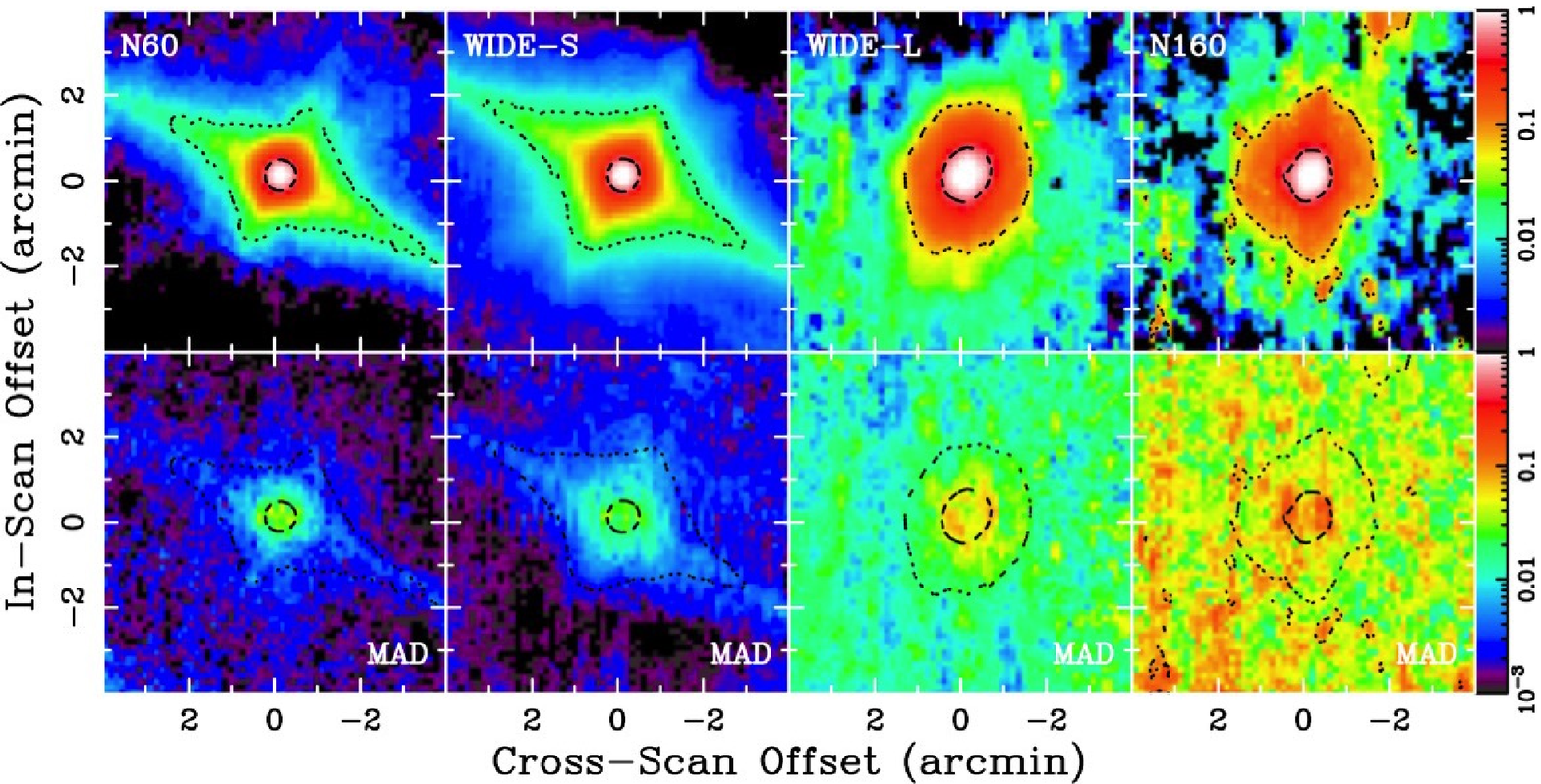}
\end{center}
\caption{%
The AKARI/FIS super-PSF images (top row) and the corresponding median absolute deviation (MAD) maps (bottom row) in the N60, WIDE-S, WIDE-L,and N160 bands (top row; from left to right). The logarithmic color scaling of the images, from 0.1\,\% to 100\,\% relative to the peak intensity, is indicated in the wedge on the right. The dashed and dotted contours in the PSF surface brightness distribution images represent the FWHM and 5\,$\sigma$ levels, respectively.}\label{F:psf}
\end{figure}

Fig.\,\ref{F:psf} shows the ``super-PSF'' images of the intrinsic diffraction limit in the four AKARI/FIS bands, made by taking the median of the normalized FAST-processed FIS maps of the PSF/photometric reference sources (using 24 and 18 sources for the SW and LW bands, respectively; Table 1 of \cite{ueta2017}).
The median MADs intrinsic to the source emission are found to be
$0.2 \pm 0.1$\,\%, 
$0.3 \pm 0.2$\,\%, 
$2.1 \pm 1.1$\,\%, 
and 
$4.1 \pm 2.1$\,\%, 
for the N60, WIDE-S, WIDE-L,and N160 bands, respectively,
within the 5-$\sigma$ region (dashed contour in the PSF images),
almost identical to the previous cases \citep{ueta2017}.

\subsection{Determination of the Updated Scaling Factors}

\begin{figure}
\begin{center}
    \includegraphics[width=\columnwidth]{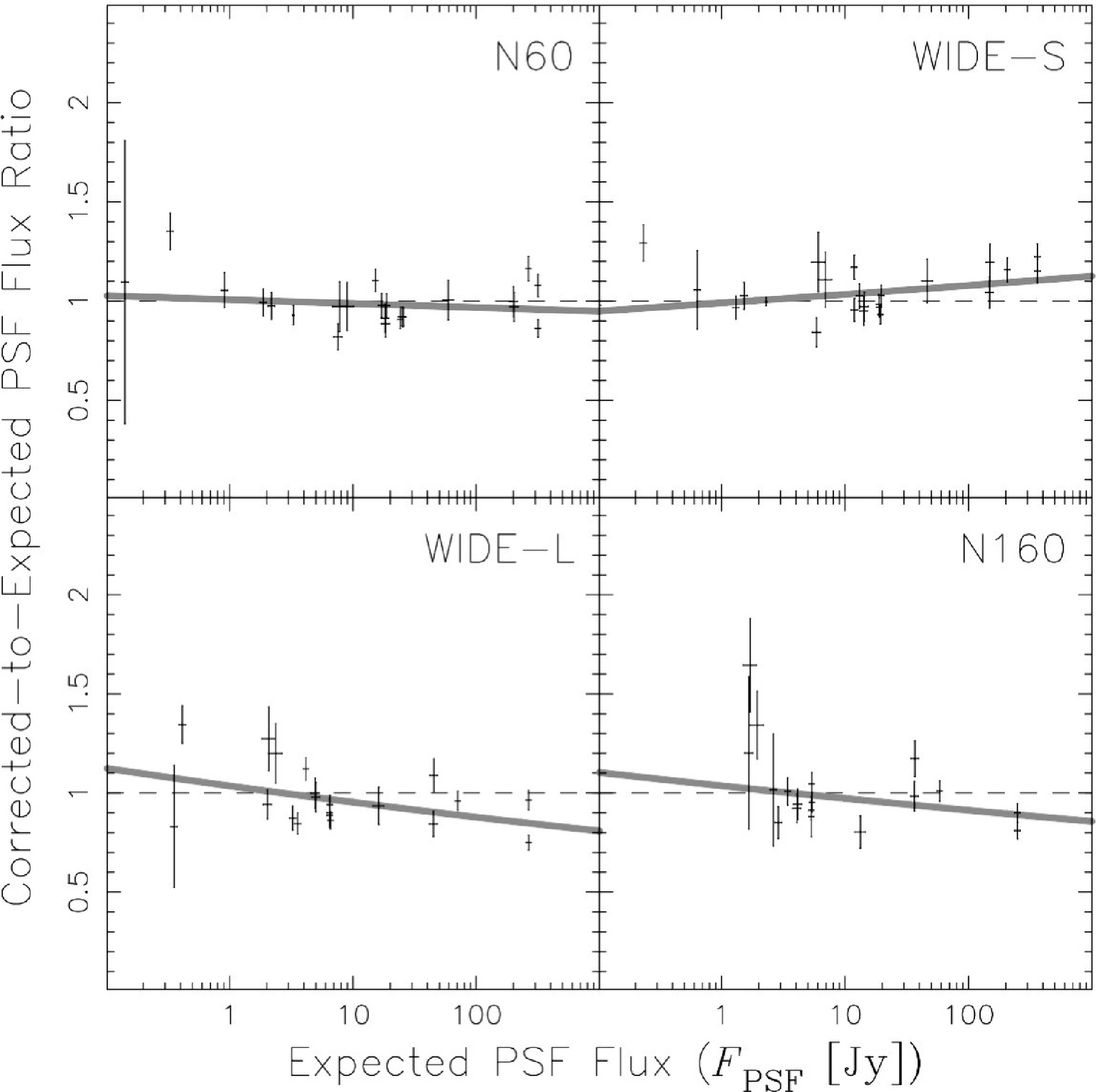}
\end{center}
\caption{%
The corrected-to-expected PSF flux ratios as a function of the expected PSF flux in each of the AKARI/FIS bands. 
The expected fluxes are the previously known fluxes of the far-IR flux calibration objects used in the analysis, while the corrected fluxes are those measured from the AKARI/FIS maps rescaled with the parameters in Table\,\ref{nandcvalues}. 
The gray solid lines show the power-law best fits. 
These plots,
compared with Fig.\,5 of \citet{ueta2017}, confirm that the revised surface brightness correction applied to the raw/archived FIS maps suppressed the signal-strength-dependent response of the FIS detector in the FAST-processed FIS maps better.}\label{F:fluxcomp}
\end{figure}

We follow exactly the same procedure as described by \citet{ueta2017} in deriving the power-law FIS response function $n$ and $c$ scaling factors.
In a nutshell, we fine-tune the $n$ and $c$ parameters until the resulting ``corrected" measured fluxes reproduce the ``expected" values of far-IR photometric calibration objects \citep{shirahata2009,ueta2017}.
The derived parameters are shown in Table\,\ref{nandcvalues}.
Here, we just present the comparison ratio between the corrected and expected PSF fluxes as a function of the expected PSF flux for each FIS band in Fig.\,\ref{F:fluxcomp}.
The power-law fits of the corrected-to-expected PSF flux ratios are 
$(1.01 \pm 0.03) \times F_{\rm Jy}^{(-0.009 \pm 0.007)}$,
$(0.99 \pm 0.02) \times F_{\rm Jy}^{(0.019 \pm 0.006)}$,
$(1.04 \pm 0.03) \times F_{\rm Jy}^{(-0.036 \pm 0.008)}$, and 
$(1.04 \pm 0.03) \times F_{\rm Jy}^{(-0.027 \pm 0.010)}$, 
respectively, for the N60, WIDE-S, WIDE-L, and N160 bands, where $F_{\rm Jy}$ is the 3\,$\sigma$ expected flux in Jy.
Compared with the previous case, the present re-scaling is a few percent better in the SW band and equally good in the LW band.
This suggests that the quality of the results of the flux calibration does not differ much by the choice of the PSF shape (i.e., the GCF parameters).

%%%
% Bibliography
%%%

%\begin{thebibliography}{}
%% Journals(e.g. A\&A,ApJ,AJ,NMRAS,PASP ...)
%% Authors, Year, Journal, Vol#, Page#
%% Journal Title Abbreviation >> http://www.asj.or.jp/pasj/Jabb.html
%\bibitem[Aauthor et al.(2001)]{key-1}
%  Aauthor, A., Bauthor, B., Cauthor, C.\ 2001, PASJ, vol, page
%\bibitem[Aauthor \& Bauthor(2003a)]{key-2}
%  Aauthor, A., \& Bauthor, B.\ 2003a, PASJ, vol, page   
%\bibitem[Aauthor \& Bauthor(2003b)]{key-3}
%  Aauthor, A., \& Bauthor, B.\ 2003b, PASJ, vol, page  
%\bibitem[Aauthor, Cauthor, and Dauthor(2000)]{key-3}
%  Aauthor, A., Cauthor, C., \& Dauthor, D.\ 2000, PASJ, vol, page   
%% Books
%\bibitem[Aauthor \& Eauthor(2003b)]{key-3}
%  Aauthor, A., \& Euthor, E.\ 2003b, Name of Book (Tokyo: Publisher) ch.0    
%% Editorial Books
%\bibitem[Dauthor(2001)]{key-n}
%  Dauthor A.~A.\ 2001, in Name of Book,
%   ed.\  D.~Editor (Tokyo: Publisher), page
%\end{thebibliography}

\begin{landscape} 
\setlength{\topmargin}{40mm}
\setlength{\tabcolsep}{4pt}
\setlength{\textheight}{160mm}
   %\begin{landscape}
\scriptsize
\begin{longtable}{llllccccccc}
%\begin{longtable}{p{2.2cm}p{1.7cm}p{1.6cm}p{1.7cm}p{0.6cm}p{1.2cm}p{1.9cm}p{1cm}p{1cm}p{1.4cm}p{1cm}}
%\begin{longtable}{*{11}{l}}
\caption{Target List and Observation Log}\label{tab:targets}
\hline
Target Name\footnotemark[a] & 
Alt.\,Name & 
RA(J2000) & 
DEC(J2000) & 
Type\footnotemark[b] & 
Var.\footnotemark[c] & 
Sp.\,Type & 
Obs.\ Date  &
Obs.\ Time & 
OBSID &
AOTPARAM\footnotemark[d]\\
\hline
\endfirsthead
\hline
Target Name\footnotemark[a] & 
Alt.\,Name & 
RA(J2000) & 
DEC(J2000) & 
Type\footnotemark[b] & 
Var.\footnotemark[c] & 
Sp.\,Type & 
Obs.\ Date  &
Obs.\ Time & 
OBSID &
AOTPARAM\footnotemark[d]\\
\hline
\endhead
\hline
\endfoot
\multicolumn{11}{l}{\parbox{6in}{%
a: IRAS catalog names, unless not observed by IRAS.
The leading ``R'' before the coordinate designation indicates that the entry is taken from the IRAS Point Source Reject Catalog \citep{IRASReject}, while ``F'' indicates RAS Faint Source Catalog \citep{IRASFaint}.\\
b: Simbad Object Type: 
AB$\ast$ (AGB star), 
C$\ast$ (carbon star),
LP$\ast$ (long period variable),
Mi$\ast$ (Mira),
No$\ast$ (nova), 
OH$\ast$ (OH/IR star), 
pA$\ast$ (post-AGB star),
PN (planetary nebula),
RC$\ast$ (R CrB star),
S$\ast$ (S star),
s$\ast$r (supergiant star),
Sy$\ast$ (symbiotic star),
and
SNI (type I supernova).
``?'' indicates that the type identification is tentative (i.e., a candidate).\\
c: Variability Type: 
``:'' indicates that the type identification is tentative (i.e., a candidate).\\
d: Parameters for the FIS01 AOT \citep{kawada2007}: (reset interval [s]);(scan speed [arcsec\,s$^{-1}$]);(shift size [arcsec]).}}\\
\endlastfoot
IRAS 00042$+$4248 	&	KU And                 & 00:06:52.752 & $+$43:05:02.25 & Mi$\ast$ & M & M10              & 7/14/07 & 19:41:45 & 1711042-001 & 1.0;15;70 \\
IRAS 00102$+$7214  & NGC 40                    & 00:13:01.01014 & $+$72:31:19.0944 & PN  & $\cdots$ & [WC8]            & 8/15/07 & 17:39:45 & 1711293-001 & 1.0;15;70 \\
IRAS 00128$-$3219  & S Scl                  & 00:15:22.26580 & $-$32:02:42.9899 & Mi$\ast$ & M & M7/8IIIe         & 6/10/07 & 18:50:25 & 1711091-001 & 1.0;15;70 \\
IRAS 00245$-$0652  & UY Cet                 & 00:27:06.4494 & $-$06:36:16.871 & AB$\ast$ & SRB & M7               & 6/25/07 & 21:42:22 & 1711023-001 & 1.0;15;70 \\
IRAS 00248$+$3518  & AQ And                 & 00:27:31.68373 & $+$35:35:14.5034 & C$\ast$  & SR & C                & 7/14/07 & 21:23:34 & 1711226-001 & 1.0;15;70 \\
IRAS 00254$-$3317  & $\eta$ Scl                 & 00:27:55.69870 & $-$33:00:25.7977 & LP$\ast$ & LB & M2/3III          & 6/12/07 & 13:56:21 & 1711122-001 & 2.0;8;70 \\
IRAS 01037$+$1219  & IRC $+$10011                & 01:06:25.984 & $+$12:35:53.05 & OH$\ast$ & M & M8               & 7/12/07 & 14:47:50 & 1711040-001 & 0.5;15;70 \\
IRAS 01144$+$6658  & RAFGL 190                 & 01:17:51.378 & $+$67:13:53.34 & C$\ast$  & M & $\cdots$ & 8/15/07 & 21:00:53 & 1711305-001 & 1.0;15;70 \\
\hline
\end{longtable}
%\end{landscape}
\end{landscape}

\begin{table}
  \tbl{Parameters and Characteristics of FIS Surface Brightness Correction}{%
  \begin{tabular}{lcccr}
      \hline
 \multicolumn{1}{l}{} & \multicolumn{2}{c}{Power-Law Fit Parameters} & \multicolumn{1}{c}{Correction Accuracy} & \multicolumn{1}{c}{Flux Range} \\
\cline{2-3}
\multicolumn{1}{l}{Band} & \multicolumn{1}{c}{n} & \multicolumn{1}{c}{c} & \multicolumn{1}{c}{(\%)} & \multicolumn{1}{c}{(Jy)}\\
      \hline
N60     & 0.897$\pm$0.009 & 1.23\phantom{0}$\pm$0.03\phantom{0} & \phantom{1}97$\pm$11 &  0.14 -- 320\\
WIDE-S  & 0.875$\pm$0.009 & 1.30\phantom{0}$\pm$0.02\phantom{0} & 103$\pm$11 &  0.23 -- 360\\
WIDE-L  & 0.949$\pm$0.009 & 1.16\phantom{0}$\pm$0.02\phantom{0} & \phantom{1}94$\pm$16 &  0.36 -- 270\\
N160    & 0.977$\pm$0.010 & 0.462$\pm$0.012 & \phantom{1}97$\pm$21 &  1.7\phantom{1} -- 250\\
      \hline
    \end{tabular}}\label{nandcvalues}
\begin{tabnote}
Note: The derived $(n,c)$ parameters are valid only when the surface brightness units of the input {\sl AKARI\/} slow-scan maps are given in MJy\,sr$^{-1}$.
\end{tabnote}
\end{table}

\scriptsize
%
 
\end{landscape}

%\clearpage

%\setcounter{figure}{0} 
%\input{supplementary_figures.tex}

\end{document}